\documentclass[prl,twocolumn,showpacs,floatfix]{revtex4}
\usepackage{graphicx}

\begin{document}
\title{Recursiveness, Switching, and Fluctuations in
a Replicating Catalytic Network}
\author{Kunihiko Kaneko\\
{\small \sl Department of Basic Science,
College of Arts and Sciences,
University of Tokyo,
Komaba, Meguro-ku, Tokyo 153, Japan}\\
}

\date{}

\begin{abstract}
A protocell model consisting of mutually catalyzing molecules is studied in order to investigate
how chemical compositions are transferred recursively through cell divisions 
under replication errors. Depending on the path rate, the numbers of molecules and species, 
three phases are found: fast switching state without recursive production, recursive production,
and itinerancy between the above two states. 
The number distributions of the molecules in the recursive states are shown to be 
log-normal  except for those species that form a core hypercycle, and are explained with the help
of a heuristic argument.
\end{abstract}
%\vspace{.4in}

\maketitle
%\pagebreak
%\section{Introduction}

In a cell, a huge number of chemicals is synthesized by mutual catalyzation leading to
replication of molecules that allow a cell to grow until it is large enough to divide into two.
How the underlying reaction networks give rise to the recursive production of cells
is an important question, not only when  considering the origin of life \cite{Eigen,Dyson} but also
when trying to understand the general features of a modern cell's biochemical reaction dynamics.

As a simple prototype of a reproducing cell,
let us consider a set of chemicals with some catalytic activities.
How can such a system consisting of chemicals connected by a 
catalytic reaction network sustain
recursive production?  Are there any generic properties
in the dynamics and fluctuations of such reproducing systems?

These questions were originally addressed
in connection with the origin of life.  Eigen and Schuster proposed the
hypercycle\cite{Eigen} as a mechanism to overcome an inevitable loss in the catalytic activities
through mutations, while Dyson\cite{Dyson} argued that it is possible for 
a collection of chemicals to be sustained by mutual catalytic activity.
Although the hypercycle itself may be susceptible to destruction by parasitic
molecules, i.e., molecules which are catalyzed by the hypercycle species
but themselves do not catalyze other molecules, it was later shown that
compartmentalization by a cell structure or localized patterns
in reaction-diffusion systems may suppress the invasion of
parasitic molecules \cite{Hogeweg,McCaskill}.

%Depending on the role of molecules
%within the catalytic reaction network, it is shown that the
%distirbution of molecule numbers (by cells) has a different form.  
%Molecules in the core part
%of catalytic reaction network have small number flucutuations, while
%at the peripheral part in the networks,  the number distributions show
%log-nromal distribution, and the parsistic molecules without recursive production shows
%a power-law distribution.  Distributions of gene expressions or some molecule concentrations
%accroding to flow-cytometry expereiments are discussed from this veiwpoint.

%\section{Model}

Here, we study a simple model of mutually catalyzing molecules and
classify the biochemical states according to their ability for recursive reproduction.
Besides fixed recursive states, we find fast switching states and 
  several quasi-recursive 
states that allow for both recursive reproduction and evolution.
Last, we study the characteristics of the number distributions of the molecular species
in these replicating cells.

We envision
a (proto)cell containing $k$ molecular species with some of the species possibly having 
a zero population. 
A  chemical species can catalyze the synthesis of some other chemical species as
\begin{equation}
[i] + [j] \rightarrow [i] + 2[j] .
\label{reaction}
\end{equation}

\noindent
with $i,j=1,\cdots,k$ according to
a randomly chosen reaction network (with a  connection rate of the
catalytic path given by $\rho$)  which is kept fixed throughout each simulation.
%(which will be taken to be 0.2 or 0.1 for most simulations here), 
%and the connection is chosen randomly. 
%Considering catalytic reaction dynamics, the reverse reaction
%process is neglected, and reactions $i \leftrightarrow j$ are not included.
%(Here we investigated the case without direct mutual
%connections, i.e.,  $i\rightarrow j$ was excluded as a possibility when there 
%was a path $j \rightarrow i$, although this condition is not essential for the 
%results to be discussed). 
Furthermore, each molecular species $i$ has a randomly chosen
catalytic ability $c_i \in [0,1]$.
(I.e., the above reaction (\ref{reaction}) occurs with the rate $c_i$). 
Assuming an environment with an ample supply of chemicals available to the cell, 
the molecules then replicate 
leading to an increase in their numbers within a cell.  
It is the dynamics of these molecule numbers $N_i$ of the species $i$ under replication
that are our main concern here.

During the replication process, structural changes, e.g.\
the alternation of a sequence in a polymer,
may occur that alter the catalytic activities of the molecules. 
%Therefore, the activities of the replicated molecule species can differ from those of the mother species.  
The rate of such structural changes is given by the replication 'error rate' $\mu$.
As a simplest case, we assume that  this `error' leads to all other  molecule
species with equal probability, (i.e., with the rate $\mu /(k-1)$).
% and could thus regard it as a background fluctuation.
In reality of course, even after a structural change, the replicated molecule
will keep some similarity with 
the original molecule, and this equal rate of transition to other molecule species
is a drastic simplification.
%a  replicated species with the `error' would be within a limited class of molecule species.
We therefore carried out also some simulations where the  errors in replication
only lead to a limited range of molecule species and found that the simplification does not
affect the basic conclusions presented here.
%Hence we use the simplest case for most simulations.

The model is simulated as follows:
At each step, a pair of molecules, say,
$i$ and $j$, is chosen randomly.  If there is a reaction path between species $i$ and $j$,
and $i$ ($j$) catalyzes $j$ ($i$),
one molecule of the species $j$ ($i$) is added
with  probability $c_i$ ($c_j$), respectively.
The molecule is then changed to another randomly chosen species with the probability of
the replication error rate $\mu$.
When the total number
of molecules exceeds a given threshold (denoted as $N$), the cell 
divides into two such that each daughter cell inherits half ($N/2$) of the molecules of 
the mother cell, chosen randomly. 
In order to take the importance of the discreteness in the moluclue numbers \cite{Togashi} into account,
we adopted a stochastic rather than the usual differential equations approach.

The cell state at the $n$-th division  is characterized by
the molecule numbers of the chemical species $\{ N^n_1,N^n_2,\cdots ,N^n_k \}$ (with $\sum_j N^n_j =N$), while
there are four basic parameters; $N$, $k$,  $\mu$, and $\rho$.
By investigating the dynamics of one thousand randomly chosen networks, and changing the four parameters,
we have found that the behaviors of the system can be  classified into just the following three
types:

(A) Fast switching states without recursiveness

(B) Fixed recursive states

(C) Itinerancy over several quasi-recursive states

In phase (A), even though each generation has some dominating species as with regards to the 
molecule numbers, the dominating species change every few generations and information regarding
the previously dominating species is totally lost often to the point that its population drops
to zero.
%Here no stable mutual catalytic reltionships are formed.
Indeed, by autocatalytic reactions, the population of one dominant species can be amplified, but soon
it is replaced by another chemical that is catalyzed by it (see Fig.1a).

\begin{figure}
\noindent
%\vspace{-.1in}
%\hspace{-.3in}
%\epsfig{file=fig1.ps,width=.6\textwidth}
%\epsfig{file=fig2.ps,width=.6\textwidth}
%\epsfig{file=fig3.ps,width=.6\textwidth}
\includegraphics[width=57mm]{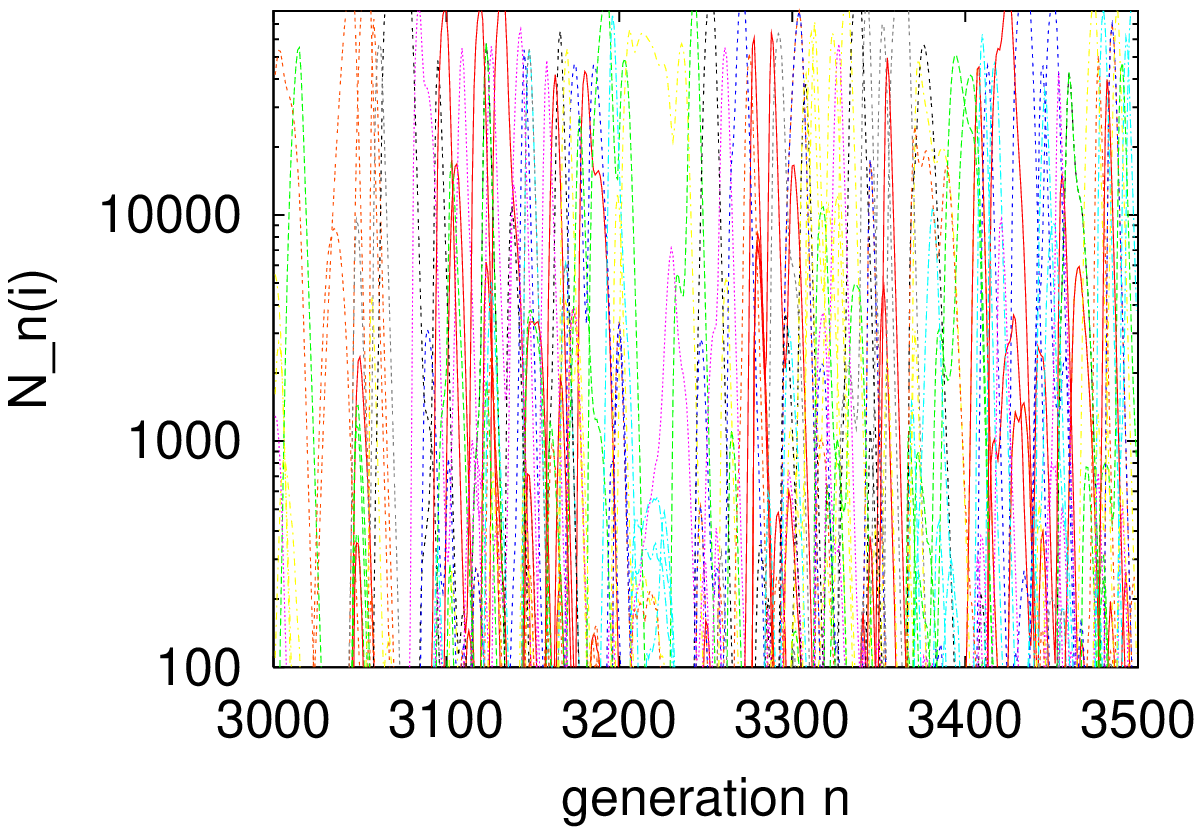}
\includegraphics[width=57mm]{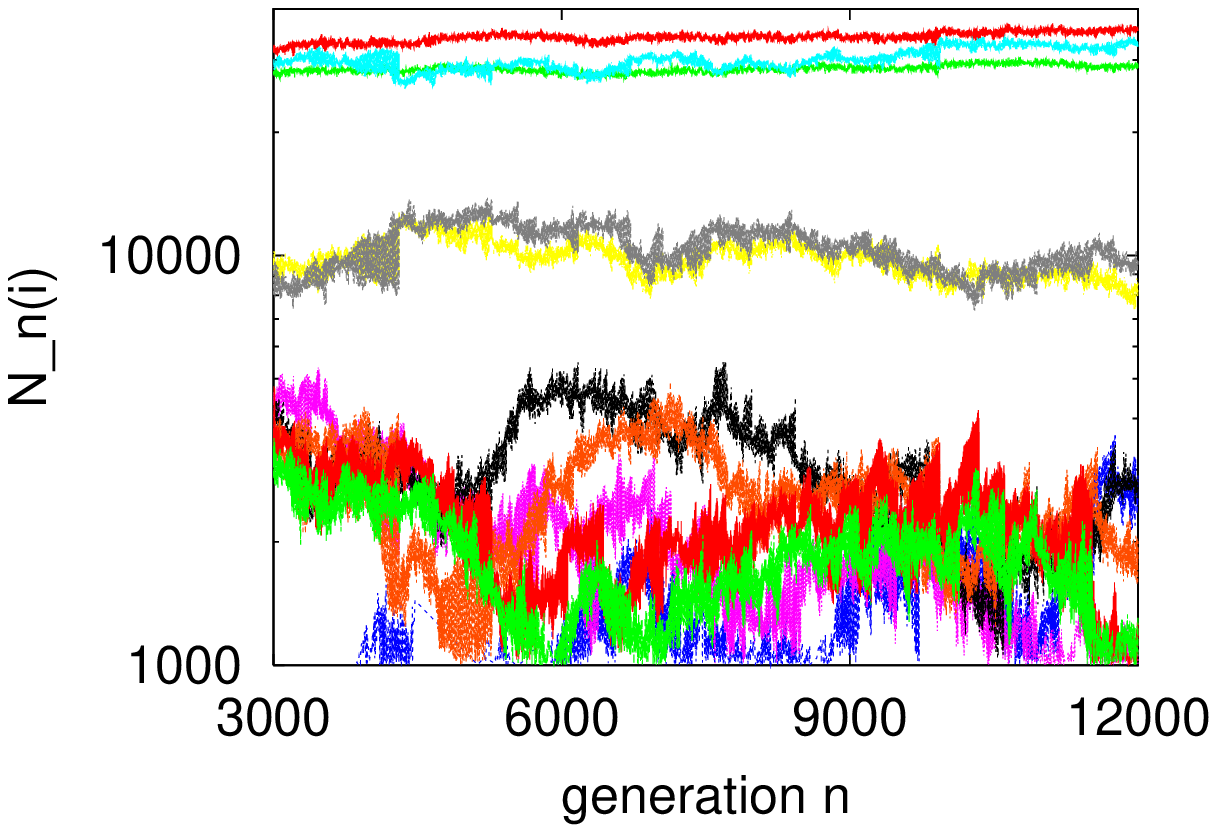}
\includegraphics[width=57mm]{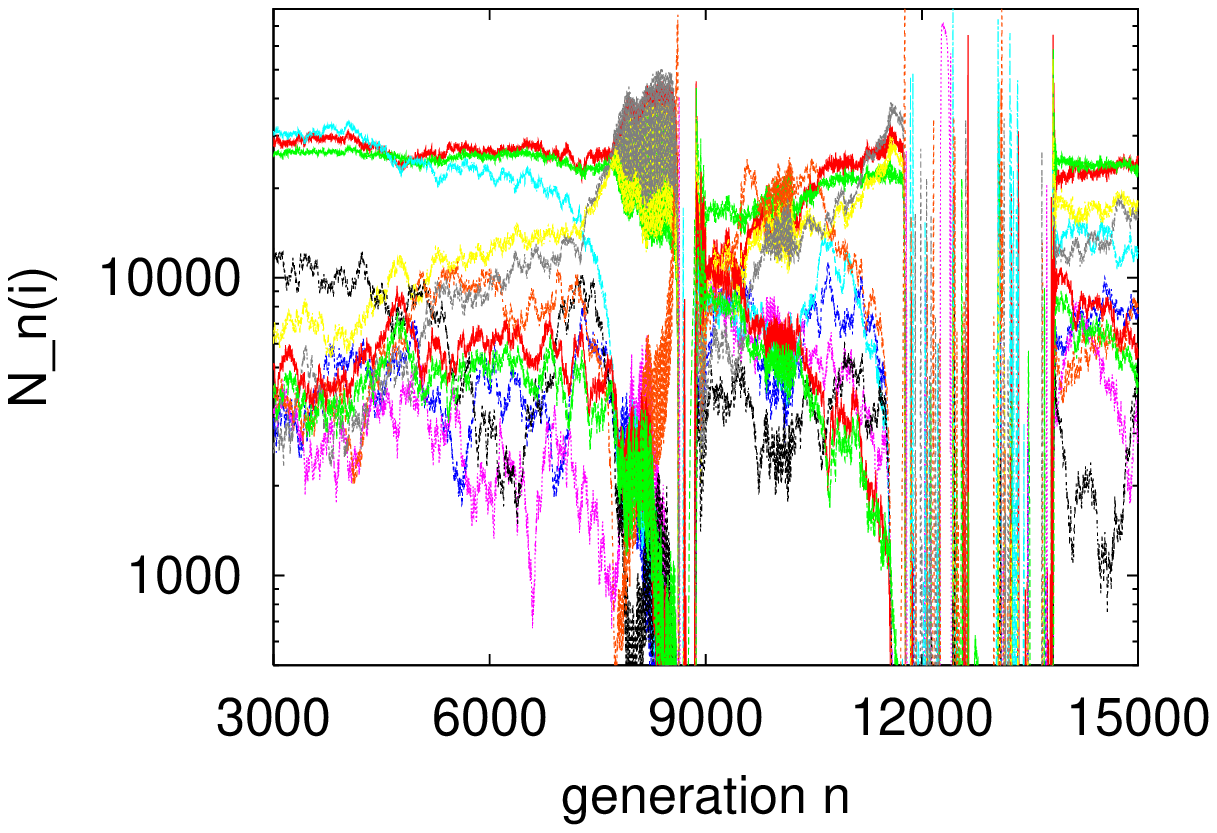}
\caption{
The number of molecules $N^n_i$ for the
species $i$ is plotted as a function of the generation $n$,
i.e., at each successive division event $n$. $\rho=0.1$, and $N=64000$.
(a) $\mu=0.01$, and $k=500$ (b) $\mu=0.01$, and $k=200$ (c)
$\mu=0.1$, and $k=200$. For (b) and (c), the same network is adopted.
Different colors correspond to different species, while 
only some species (whose population becomes large during some generation)
are plotted.}
\end{figure}

\begin{figure}
\noindent
%\hspace{-.3in}
%\epsfig{file=fig4.ps,width=.5\textwidth}
%\includegraphics[width=58mm]{fig4.ps}
\includegraphics[width=38mm]{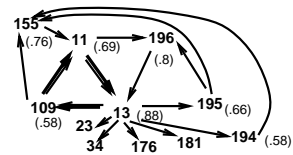}
\caption{The catalytic network of the species
that constitute the recursive state}
\end{figure}

In phase (B), a recursive state is established where
the chemical composition is stable enough to withstand the
division process. Once reached, this state lasts very long
(e.g., for as long the simulation lasts, say $10^4$ generations) (see Fig.1b).

The recursive state (`attractor') here is not necessarily a fixed point (with fluctuations)
%as with regards to the population dynamics of the chemical concentrations.s
since the molecule numbers may oscillate in time.  
Nevertheless, the overall chemical compositions
remain within certain ranges: for example, the major species
(i.e. those that are synthesized by themselves, not by an error in the replication process)
are not altered over the generations.  Generally, all the observed recursive states consist of
5-12 species, except for those species which exist only as a result of replication errors.
(see also \cite{Lancet} for recursive transmission of state in a
network model with some structure).
%where the number of molecules is one or two. 

For example, in the recursive state depicted in Fig.1b, there are 11 species
whose populations remain in existence throughout the simulation. As is shown in Fig.2,
the replication of the molecules
is sustained by the 'core hypercycle'  $109\rightarrow 11 \rightarrow 13 \rightarrow 109$
where the catalytic activities of these core species satisfy  $c_{13}>c_{109}>c_{11}$,
and accordingly we have for the respective populations 
$N_{11} > N_{109} > N_{13}$.  
This relationship is natural, since  molecules with
higher catalytic activities  result in the synthesis of more
molecules other than themselves
thus suppressing their own  population 
fraction\cite{estimate}.  

Here, through mutual catalyzation,
molecules with  higher catalytic activities are catalyzed by molecules with
lower activities but  larger populations.  
To destroy such a network of mutual support, large fluctuations in 
molecule numbers are required, which are rare for large $N$.
Hence parasitic molecule species cannot easily invade the core.

In phase (C), the system alternates between 
quasi-recursive states similar to phase (B) that last for many generations
and fast switching states similar to phase (A). 
The quasi-recursive state itself can be subjected to switches between core hypercylces
as can be seen in
Fig.1c where a switch occurs from an initial core hypercycle ($109$,$ 11$,$13$),
to the next core hypercycle $(11,13,195,155)$ around the 8500th generation. 
Subsequently, around the 12000th generation, the core network is taken over by parasites to 
enter the phase (A) like fast switching state which in turn gives way for a new
quasi-recursive state around the 14000th generation.
%The former is the compeition among core networks, while 
%the latter is invasion of parasitic molecules.
%These two types of switches are generally observed, while the latter,
%which destroys the quasi-stable recursive state, is more frequent,
%Once this occurs,
%successive take-over of parasitic molecules occurs, and 
%the fast switching state (A) appears.
% with the decrease in chemical diversity.

When $N$ is not so large, the molecule number of the 
species with the highest catalytic activity in the core hypercycle can become small
due to fluctuations, and subsequently succumb to parasitic molecules.  
%(Note the species catalyzes other molecules better, and the number is smaller.) If that happens,
Then, the core hypercycle loses its 
main catalyst resulting in its collapse giving way to a fast switching state that in
turn will allow the formation of a new core hypercycle (which can but does not have to
be identical to the previous one).

Which one of the phases (A), (B), (C) appears, of course,  depends on the parameters and the 
specific structure of the network.
There is however, a clear dependence of the fraction of the networks leading to each of 
these phases on the parameter values.
The fraction of (B) increases and the fraction of (A) decreases for
 increasing $N$, or for decreasing $k$, $\rho$ or $\mu$.
For a more systematic investigation, it is useful to classify the phases by the similarity
of the chemical compositions between two cell division events\cite{Lancet}. This can be
done
by defining a $k$-dimensional vector $\stackrel{\rightarrow}{V_n}$=$(p_n(1),..,p_n(k))$
with $p_n(i) =N_n(i)/N$ and measuring the similarity between  $\ell$  successive generations 
with the help of the inner product
$H_{\ell}=\stackrel{\rightarrow}{V_n} \cdot \stackrel{\rightarrow}{V_{n+\ell}}/(|V_n||V_{n+\ell}|)$.
In Fig.3, the average similarity $\overline{H_{20}}$ and
the average division time are plotted for 50 randomly chosen reaction
networks as a function of the path probability $\rho$.  
For $\rho >0.2$,  phase (A) is
 observed for nearly all the networks (e.g. $48/50$), while for lower path rates, the fraction of (C)
(with a few (B)) increases.
(Roughly speaking the networks  with $\overline{H_{20}}>.9$ belong to $C$, and those with $\overline{H_{20}}<.4$ to $A$).
 
In general, we have found a positive correlation between
the growth speed of a cell, 
the similarity $H$, and the diversity of the molecules.
(In Fig.3, networks with larger $H$ have smaller division times).
The recursive states, established by a variety of species, maintain higher growth speeds 
since they effectively suppress parasitic molecules. 
In Fig. 3, for decreasing path rates,
the variations in the division speeds of the networks become larger,
and some networks
that reach recursive states have higher division speeds than networks with 
larger $\rho$. On the other hand, when the path rate is too low, the protocells generally
cannot grow since the probability to have useful connections in the network is nearly zero.
Indeed an optimal path rate seems to exist (e.g., around $.05$ for $k=200$, $N=12800$
as in Fig.3) for which some networks have rather high growth speeds. Consequently, 
in an environment that necessitates competition for growth, protocells having such
optimal networks will be more successful than protocells with sub-optimal networks.

\begin{figure}
\noindent
%\hspace{-.3in}
%\epsfig{file=fig45.ps,width=.5\textwidth}
%\includegraphics[width=63mm]{fig4.ps}
\includegraphics[width=60mm]{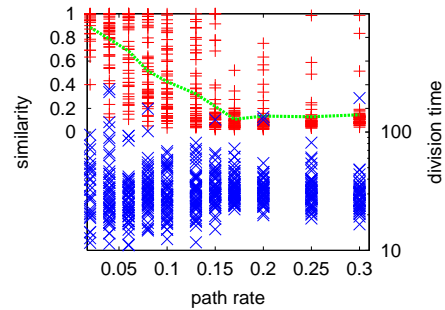}
\caption{The average similarity $\overline{H_{20}}$ (red),
and the average division time (blue) are plotted as a function of
the path rate $\rho$.  For each $\rho$, data from 50 randomly chosen networks are plotted.
The average is taken over 600 division events.  The green line indicates 
the average of $\overline{H_{20}}$ over the 50 networks for each $\rho$.
%For $\rho>.2$, networks over 98 \% have $H<.4$, and they show fast switching,
%while for $\rho=.08$, about 20??\% belong to the phase $B$, and 75??\% to $C$.
At $\rho=0.02$, 25 out of 50 networks cannot support cell growth,
4 cannot at $\rho=0.04$.}
\end{figure}

%\section{Fluctuation}
Finally, we investigate the fluctuations of the molecule numbers of each of the species.
Since the number of molecules is not very large,  
the fluctuations over the generations can possibly have a significant impact on the
dynamics of the system.  
In order to quantify the sizes of these fluctuations, have  measured the distribution $P(N_i)$ for
each molecule species $i$, by sampling over division events. 
Our numerical results are summarized as follows:

(I) For the fast switching states, the distribution $P(N_i)$ satisfies the power law
%\begin{equation}
$P(N_i) \approx N_i^{-\alpha}$, with $\alpha \approx 2$.
%\end{equation}

(II) For recursive states, the fluctuations in the core network (i.e., 
13,11,109 in Fig.2) are typically small.
% and are roughly fit by Gaussain distribution.  
On the other hand, species that are peripheral to but catalyzed by the core hypercycle have log-normal
distributions 
\begin{math} P(N_i) \approx exp(-\frac{(logN_i-\overline{log N_i})^2}{2\sigma})
\end{math}, as shown in Fig.4.
We have also plotted the variance $\overline{(N_i-\overline{N_i})^2}$ ($\overline{..}$ 
is the average of the distribution $P(N_i)$) and the deviation
between the peak of $P(N_i)$ and $\overline{N_i}$, divided by the average $\overline{N_i}$. 
As can be seen, the variance and the fluctuations in the core network are small,
especially for the minority species (i.e., 13). 
For molecule species that do not belong to the core hypercycle,  the variance scaled
by the average increases as the average decreases.  Furthermore, there
is a distinct deviation between the peak and the average (except for the core
species),  since the distribution
has a tail for larger sizes.  On the other hand, if we 
use the variable $log N_i$ when plotting the distribution, it 
is closer to a Gaussian, and the difference between the peak
and the average is suppressed.

The origin of the log-normal distributions here can be understood by
the following rough argument: for a replicating
system, the growth of the molecule number $N_m$ of the species $m$ is given by
\begin{math}
dN_m/dt=AN_m
\end{math}
where $A$ is the average effect of all the molecules that catalyze $m$.
%
% NOTE: This sounds kind of complicated, perhaps an inline formula would be better.
%
We can then obtain the estimate 
\begin{math}
dlog N_m/dt =\overline{a} +\eta(t).
\end{math}
by replacing $A$ with its temporal average $\overline{a}$ plus fluctuations $\eta(t)$ around it
\cite{estim2}.
If $\eta(t)$ is approximated by a Gaussian noise,
the log-normal distribution for $P(N_m)$ is suggested
(this argument is valid if $\overline{a}>0$). 
For the fast stitching state
the growth of each molecule species is close to zero on the average and  
in this case, by considering the Langevin equation with boundary conditions, the power law follows
as discussed in \cite{Sornette,Mikhailov}.  

\begin{figure}
\noindent
%\hspace{-.3in}
%\epsfig{file=fig4.ps,width=.5\textwidth}
\includegraphics[width=58mm]{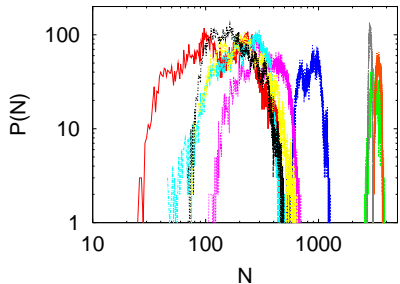}
\caption{The number distribution of the molecules corresponding to the network in Fig.2.
The distribution is sampled from 1000 division events.  From right to left, the plotted 
species are 11,109,13,155,176,181,195,196,23.
Log-Log plot.}
\end{figure}

\begin{figure}
\noindent
%\hspace{-.3in}
%\epsfig{file=fig5.ps,width=.5\textwidth}
%
% NOTE: it is quite difficult to see the data of this figure.
%
%
\includegraphics[width=58mm]{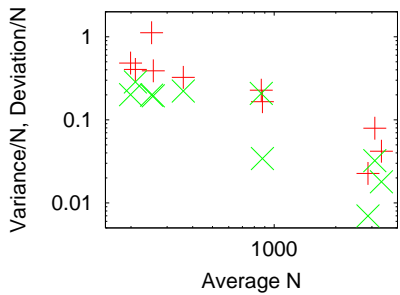}
\caption{Scaled variance and deviation.  $\times$ denotes the variance of the molecule number
divided by its average, while $+$ shows the difference between the average and the peak 
of the distribution divided by the average $\overline{N_i}$.  
From the right to left, the species
11,109,13,155,194,176,195,181,196,23,34 are plotted, i.e.,
the largest $\overline{N_i}$ corresponds to species 11, the third largest to 13, 
and so forth. Computed in the same way as Fig.4. }
%The latter is positive since the distribution $P(N)$ has
%a tail to larger value, while by taking $log N$, the deviation is decreased drastically
%for the  species except the hypercycle core three species, 11,109, and 13 that have
%three largest average $\overline{N}$.  Note small variance of the species 13 (minority in the
%hypercycle). Computed in the same way as Fig.4. }
\end{figure}

If several molecules mutually catalyze each other, due to the central limit theorem,
one would expect their distributions to be close to Gaussian, and this is indeed the
case for the three core species.

%For some networks, the distributions of the molecule numbers in
%the recursive sates may sometimes be intermediate between log-normal and Gaussian, and
%occaisionally even have double peaks.
By studying a variety of networks, the observed distributions of the molecule numbers
can be generally summarized as:
(1)Distribution close to Gaussian form,
with relatively small variances in the core (hypercycle) of the network.
(2)Distribution close to log-normal, with larger fluctuations
for a peripheral parts of the network. 
(3) Power-law distributions
for parasitic molecules that appear intermittently\cite{rem}.

%
% NOTE: By itself the following sentence seems to convey little meaning.
%
%
%It is interesting to examine the distributions of several chemicals in cells, to classify the
%istribution form, in relationship with biochemical reaction networks. 
%
%Using recent advances in  quantatiataive measurements of the flucutuations,
%study of distribution forms will be basically important
%and the relationship of the distribution form and the role of molecues
%in the catalytic networks should be searched for.
%
%
% NOTE: I think the strength of the conclusion would benefit from making your points a 
% bit more explicit... if there's enough room
%
%

To sum up, features of a protocell with  
catalytic reactions and divisions are classified into three phases.  Recursive  
states (B) and switching over quasi-recursive states (C) should be noted,
that maintain catalytic activities for cell reproduction. 
Besides the establishment of recursive growth, 
variability of cells in their chemical compositions is necessary, in order to ensure  evolvability.
Previously, we pointed out the relevance of minority molecules 
in mutually catalyzing systems for making recursive growth and evolution 
compatible\cite{KK.MCT}.
Indeed, phase (C) 
satisfies both the features, since
novel quasi-recursive states with different chemical compositions are visited successively,
triggered by extinctions of minority molecules in the core hypercycle networks.

We showed 
suppression of the fluctuation of molecules at a core hypercycle
network and ubiquity of log-normal distribution of those at a peripheral network,
which can be testified for the present cell,
using recent advances in  quantitative measurements of the fluctuations.
Indeed, it is interesting to note that the distributions of the abundances of
fluorescent proteins, measured by flow-cytometry are often closer to
log-normal than Gaussian\cite{yomo}.  Furthermore, e.g.\, the size of bacteria\cite{yomo} and some
cells in blood\cite{size}  (as well as human body weight) obey log-normal distributions.  

%The itinerancy state (C) has potentiality for evolution, termed as evolvability, since 
%a novel state with different chemical compositions is visited successively.

%To see this potentiality, we have also studied a model with two modifications;
%first the catalytic activity ise set as $c_i=i/k$, i.e., the activity
%is monotonically increasing with the species index, and instead of global change
%to any molecules species by replication error, we set the change occurs only within a given
%range $i_0 \ll k$ (i.e., when the molecule species $j$ is synthesized, with the error rate
%$\mu$, the molecule $j+j'$ with $j'$ a random number over $[-i_0,i_0]$ is syntheized).
%Starting from species with $i<i_{ini}$, one can examine, if the evolution to have
%higher catalytic activity progresses or not.  As is expected, for the itinerancy state
%after one recursive state ( consisting of species within the width of 
%the order $2i_0$ ) and lasts for some time, a new switching occurs to increase
%the catalytic activity (i.e., the index of species $i$).   Once the
%species with the highets catalytic activity, and minority in the core decreases its population,
%switching to a new state that has a higher catalytic activities, and the 
%species indices successively increase.  Hence, evolution from a rather primitive cell
%consisting of low calatalytic activities to a higher actitivityes is possible.

%{\bf Acknowledgment}

I would like to thank C. Furusawa, F. Willeboordse,  and  T. Yomo
for useful discussions.
%and F. Willeboordse for critical reading of the manuscript.
This research was supported by 11CE2006.
%Grants-in-Aid for Scientific Research from
%the Ministry of Education, Science and Culture of Japan (11CE2006).

%\addcontentsline{toc}{section}{References}


\begin{thebibliography}{999}
\bibitem{estimate}
The  population sizes
of just three species $A,B,C$, with the catalytic relationship $A\rightarrow B \rightarrow C \rightarrow A$ 
are estimated by taking the continuum limit $N \rightarrow \infty$ and
obtaining a fixed point solution of the rate equation for the concentrations
of the chemicals. From a straightforward calculation we have: $N_A:N_B:N_C= c_B:c_A:c_C$.
\bibitem{estim2}
As such this equation diverges with time, but here, 
the cell divides into two before the divergence becomes significant.  
The finiteness of the total number $N$ means that the asymptotic
distribution as $N \rightarrow \infty$ is not available,
but the argument on the distribution form is valid as long as $N$ is sufficiently large.
\bibitem{rem}
For some networks, the distributions 
may sometimes be intermediate between log-normal and Gaussian, and
even have double peaks.
\end{thebibliography}
\end {document}